\documentclass[lettersize,journal]{IEEEtran}
\usepackage{amsmath,amsfonts}
\usepackage{algorithmic}
\usepackage{algorithm}
\usepackage{array}
\usepackage[caption=false,font=normalsize,labelfont=sf,textfont=sf]{subfig}
\usepackage{textcomp}
\usepackage{stfloats}
\usepackage{url}
\usepackage{verbatim}
\usepackage{graphicx}
\usepackage{cite}
\usepackage{amssymb}
\usepackage{comment}
\usepackage{siunitx}
\begin{document}

\title{Spatially-resolved charge detectors for particle beam optimization with femtoampere resolution achieved by in-vacuum signal preamplification}

\author{Kilian Brenner\textsuperscript{1}, Michael Zimmermann\textsuperscript{1}, Maik Butterling\textsuperscript{2}, Andreas Wagner\textsuperscript{2}, Christoph Hugenschmidt\textsuperscript{1} and Francesco Guatieri\textsuperscript{1 \textdagger}
\thanks{\textsuperscript{1}Forschungs-Neutronenquelle Heinz Maier-Leibnitz (FRM II), Technical University of Munich Lichtenbergstr. 1 85746 Garching, Germany}%
\thanks{\textsuperscript{2} Helmholz Zentrum Dresden Rosendorf, Bautzner Landstra\ss{}e 400, 01328 Dresden, Germany}
\thanks{\textsuperscript{\textdagger} Corresponding author}}

\markboth{IEEE Transactions on Instrumentation and Measurement }%
{Shell \MakeLowercase{\textit{et al.}}: A Sample Article Using IEEEtran.cls for IEEE Journals}

\IEEEpubid{}

\maketitle

\begin{abstract}
We present the design of a Faraday cup-like charged particle detector in a four quadrant configuration aimed at facilitating the alignment of low-intensity beams of exotic particles. The device is capable of assessing the current on the electrodes with a resolution of \SI{33}{\femto A} within \SI{15}{\milli s} or a maximal resolution of \SI{1.8}{\femto A} with a measurement time of \SI{12.4}{s}. This performance is achieved by minimizing the noise through a preamplification circuit installed in vacuum, as close as possible to the electrodes. We tested the detector with the positron beam of ELBE, achieving the nominal maximum resolution with high reproducibility. We then exploited the capabilities of the detector to resolve the two-dimensional shape of the beam, and revealed the presence of a weak electron beam being transported alongside the positrons. Characterization of the detector performance showed that in a variety of scenarios it can be used to quickly center positron beams thus allowing for the prompt optimization of beam transport.
\end{abstract}

\begin{IEEEkeywords}
Beams, Detectors, Faraday cups, Positrons, Femtoscale
\end{IEEEkeywords}

\section{Introduction}
\label{sec:Intro}

\IEEEPARstart{F}{araday} cup detectors sense the presence of a beam of charged particles by collecting them on an electrode and then measuring the deposited charge~\cite{Brown_1956}. In their simplest realisation they consist of a bare electrode, while more complex designs can contain baffles and suppression electrodes to reduce the systematics introduced by secondary emissions~\cite{Mayer1,IsoldeGuardedCup}, or grids to select particles depending on their kinetic energies. Their most common application is to provide an accurate measurement of the beam intensity; although variants exist that are made to be position-sensitive by installing multiple electrodes~\cite{Papa1} or by moving them across the particle beam~\cite{NewPlasmaDiagnostics,Sosolik1}. Faraday cups can serve as detectors for particle physics experiments~\cite{Williams1,Leland1} and are routinely used as a diagnostic tool for particle beams~\cite{ColliderFC}, plasma experiments~\cite{JETFaradayCups}, ion thrusters~\cite{RecommendedFC} and even in space probes, including the Voyager spacecrafts~\cite{Voyager_1977}.

The implementation of a Faraday cup detector for a beam current that exceeds one nanoampere (\SI{1}{nA}) is fairly straightforward. Detecting a continuous beam of positrons (or other exotic charged particles such as muons~\cite{Cook2017}) can be challenging, as their current typically ranges from less than one femtoampere (\SI{1}{fA} $\simeq$ \SI{6.2e3}{e^+\per\second})~\cite{Brusa1} to at most hundreds of picoampere (\SI{100}{pA} $\simeq$ \SI{6.2e8}{e^+\per\second})~\cite{HugenschmidtPrimaryBeam,PiochaczRemoderatedBeam}.

A Faraday cup designed to detect beams of extremely low intensity requires a charge collecting amplifier which is typically installed outside of the vacuum chamber and connected to the electrode via a pass-through. This kind of detector requires typically a long integration time, often in the range of minutes, since the long ultra-high-impedance connection between the charge collecting amplifier and the electrode acts as an antenna gathering the ambient electromagnetic noise, which can be averaged out only over a long time. 

The precision of the beam current measurement is fundamentally limited by the Cramer-Rao bound~\cite{van2004detection}, stating that the variance of an (unbiased) estimator improves at most with the square-root of the integration time. Therefore,
it is impossible to increase arbitrarily the speed of a Faraday cup detector while reaching at the same time a given measurement precision, unless the electromagnetic noise picked up by the electrode is reduced by physical means.

We present here a design for a four quadrant Faraday cup-like detector aimed at performing the alignment of low-intensity charged particle beams. The detector needs to be at the same time fast enough to enable an expedient beam optimization and sensitive enough to be able to detect currents smaller than \SI{1}{\pico A}. We achieved these goals by reducing the connection length between the electrodes and amplifiers by installing preamplifiers directly into the vacuum chamber, on the rear side of the electrodes.
This design is in some ways similar to that of the ``Fast Faraday cup detector'' previously installed at ISOLDE~\cite{Focker1}, whose designs were able to achieve a bandwidth in the \SI{}{\giga Hz} range,
albeit with currents in the order of milliamperes. To our knowledge the construction of a detector with this general design but aimed at working in the sub-picoampere range has never been attempted.
Characterization of the device shows that this design meets the design goal requirements.

\section{Design of the device}
\label{sec:Design}

\subsection{Analog frontend}

\begin{figure}
    \centering
    \includegraphics[width=0.95\linewidth]{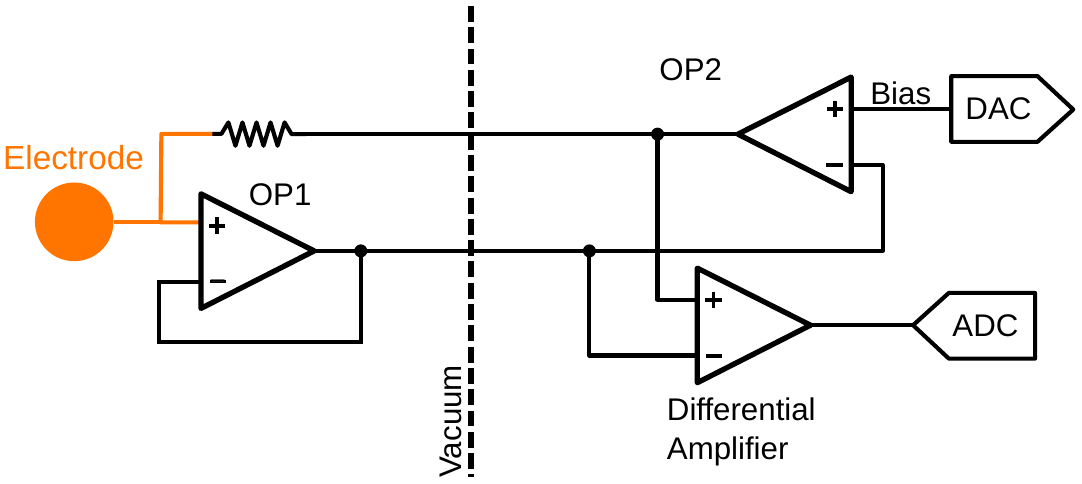}
    \caption{Simplified schematic of the main feedback loop at the center of the detector. The feedback ensures the electrode is kept at the voltage \emph{Bias}, which can be trimmed on a per-electrode basis. Since this is achieved through a high impedance resistor, the feedback voltage is proportional to the sum of the beam and leakage current on the electrode. The connections highlighted in orange are high-impedance and are physically built to be as short as reasonably achievable.}
    \label{fig:Loop}
\end{figure}

The sensor we present in this work consists of four electrodes arranged in a four-quadrant configuration to allow sensing the position of the positron beam relative to the center of the quadrant. Each electrode is read out using a feedback loop (shown simplified in figure \ref{fig:Loop}) which has three components installed inside of the vacuum chamber: a gold-plated electrode acting as the positron collector, an operational-amplifier (Op-Amp)\footnote{Analog Devices ADA4530-1} installed in voltage follower configuration and a $\SI{1.00 \pm 0.01}{\giga\ohm}$ resistor. The Op-Amp selected to fulfill this function (OP1) is optimised for the assessment of vanishingly small currents and was chosen due to its guaranteed low input leakage ($\SI{20}{\femto\ampere} \simeq  \SI{1.25e5}{e^+/\second}$). OP1 relays the voltage of the electrode within \SI{50}{\micro V} while reducing its impedance from the petaohm range to \SI{20}{\ohm}, thereby mitigating the noise across the feed-line to the outer part of the circuit by up to 14 orders of magnitude.

The lower line in the schematic is connected to a second Op-Amp\footnote{Texas Instruments TL082 or ADA4530-1, depending on board assembly} installed outside of the vacuum chamber (OP2) which keeps the voltage of the electrode constant by acting on the voltage on the other side of the gigaohm resistor. To counteract the change in voltage induced by the positron current impinging on the electrode, OP2 needs to induce an equivalent current across the gigaohm resistor, resulting in a voltage decrease of \SI{1}{\milli V} per picoampere of positron current. By measuring the difference in voltage between the output of the two Op-Amps we can thus determine the positron current on the electrode.

Maintaining a constant voltage on the electrode has multiple benefits. Firstly, the leakage current to ground and among the four electrodes is made constant, allowing its subtraction as pedestal from the measurements. Secondly, if we consider the maximum ratings of leakage current and input bias for both Op-Amps we would expect the analog frontend to measure the current on the electrode with a tolerance of \SI{70}{\femto A} or \SI{5}{\pico A} depending on the Op-Amp installed on the outside. In reality, these are maximum tolerances that are guaranteed to never be breached across different specimens of the same component and across the full allowed range of input voltages and temperatures; due to the design of our sensor the input voltages of both Op-Amps is expected to stay the same regardless of the positron current on the electrode, therefore we can expect these sources of bias to stay the same both when measuring the beam and when measuring the background. As we will see, the actual resolution of the sensor is much higher than these estimates.

The difference in the output voltage of the two Op-Amps is fed into an instrumentation differential amplifier\footnote{Analog Devices AD620} configured for a factor 10 gain and fed into a passive low-pass filter. The amplification factor 10 was chosen so that the leakage current present in the system would bring the output of the amplifier close to the center of the digitization range. Along with the primary node of the instrumentation amplifier this creates a second order low-pass with poles at \SI{340}{Hz} and \SI{800}{kHz}. The lower of the filter poles has been calibrated so that the output noise is in the order of one least significant bit (lsb) unit of the digitizer (see later) to allow for dithering. The filtering stage serves the additional purpose of raising the output impedance of the amplifier so that out-of-range values can be safely clipped to the voltage rails.

\subsection{Digitization}

The signal is then digitized using the built-in analog-to-digital converter (ADC) of a microcontroller unit (MCU)\footnote{Microchip PIC24FJ32GB002}, which is multiplexed to several pins directly on the MCU die. Integration of the ADC and the MCU allows us to keep the boards digitally silent during measurement: as the ADC data is acquired, all voltages present on the traces of the Faraday readout circuit boards are kept constant (except for the unavoidable fluctuations induced by noise). If multiple samples are to be measured, they are acquired one after the other, stored in the MCU volatile memory and then re-transmitted in bulk after the measurement has concluded. This removes a slew of potential sources of electronic noise from the system (in particular, from the power rails) that could be induced by digital communication. The bottleneck limiting the speed and accuracy of measurement of this configuration is the limited 10-bit depth of the built-in ADC. As we used a \SI{2.5}{V} voltage reference for the ADC its least significant bit corresponds to a current of \SI{0.24}{\pico A} on the electrode. This resolution can be increased by applying dithering, as we will see in section \ref{subsec:Stability}.

The MCU handling the measurement can communicate with a secondary, identical MCU unit through two galvanically insulated optocouplers, the UART protocol was chosen to implement this connection. The secondary MCU has the task of relaying communications through a USB connection with a control PC. The \SI{5}{V} power of the USB connection is used to synthesize two galvanically insulated power rails that power the primary MCU and the analog frontend; therefore no power source other than the USB connection is required. Nonetheless, since the electrodes and their entire readout system are galvanically insulated, an external high-voltage power supply can be connected to the system to bias the detector up to $\pm \SI{1}{\kilo\volt}$. Biasing the detector assembly enables the possibility to accelerate low energy beams to implant them more effectively into the electrodes.

If multiple ADC samples are to be acquired, it is necessary to first record them all and then to transmit them through the UART bridge, otherwise the design goal of not having any digital noise during measurement would be defeated. We programmed the primary MCU to be able to record on demand an arbitrary number of samples, from an arbitrary subset of the available Faraday cups, with a pace that can also be selected by the user up to \SI{20.5}{kilosamples \per \second} when measuring from as single electrode or \SI{24.5}{kilosamples \per \second} when alternating the electrodes being measured.
As the internal memory of the MCU is limited, only \SI{4}{kilosamples} can be stored before retransmission must be initiated; to partially overcome this limitation we allowed the option of requesting up to 64 consecutive 10-bit samples to be added together and stored as a single 16-bit word.

\subsection{Mechanical design}

\begin{figure*}
	\centering
	\includegraphics[width=.8\textwidth]{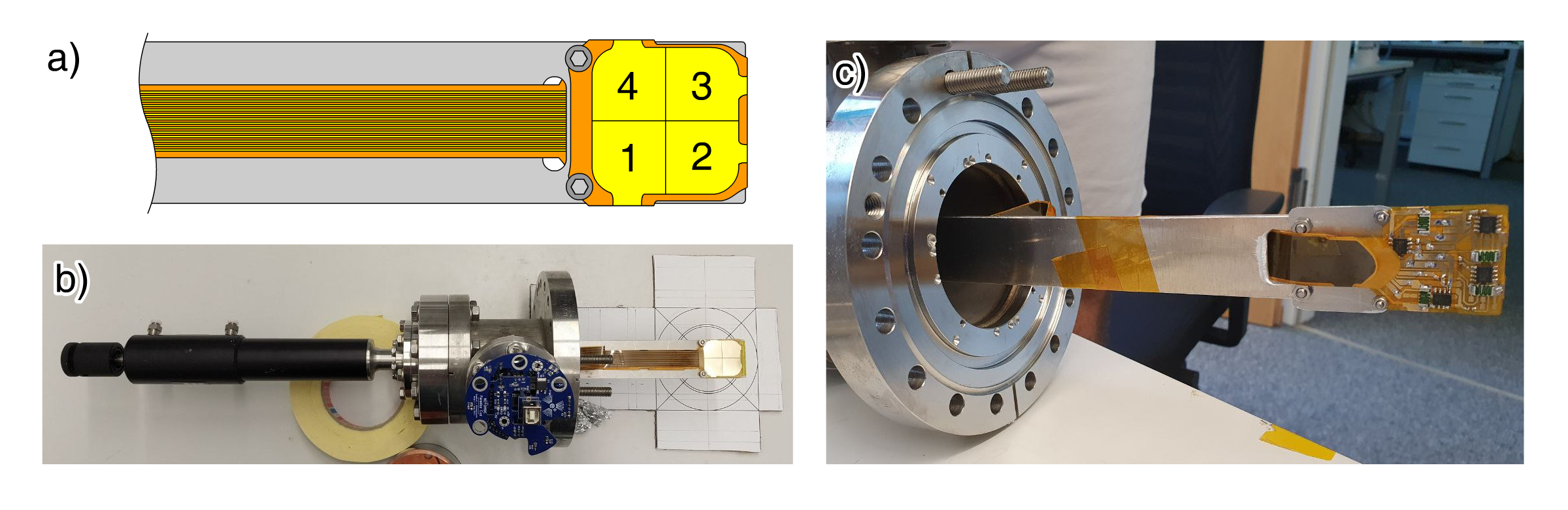}
	\caption{a) A schematic representation of the front side of the detector indicating the numeration we used to denote each of the four. b) From left to right, the manipulator assembly, the flange-mounted readout electronics and the electrode side of the Faraday cups. c) The back side of the detector is visible with the four preamplifiers installed in close proximity to the electrodes.}
	\label{fig:Assembly}
\end{figure*}

The four electrodes, the cable connecting the detector with a passthrough and the assembly substrate of the preamplifier are all realized from a single one-sided gold-finished flexible kapton\textregistered{} printed circuit board (PCB) printed without solder mask, silk screen or stiffeners to maximize vacuum compatibility (see figure~\ref{fig:Assembly}). The flexible PCB folds onto itself like an origami to produce the preamplifier assembly on the back of the electrodes, from three different flaps meeting and being soldered together. The resulting shape is easily mounted onto a manipulator, which is needed to extract the Faraday cup from the beam path after use. Due to its flexibility the PCB can be soldered directly to a commercial CF40 D-SUB15 passthrough.

The entire readout assembly is installed directly on the flange holding the passthrough. The entire system is fit onto three stacked PCBs, each having a diameter smaller than the ConFlat (CF) flange. This allows bolting the PCBs onto the flange itself through the use of threaded rods. The installation of a copper shield completely enclosing the readout assembly allows further reducing the electronic noise felt by the analog system.

\section{Characterization}

Given that the performance of the system is potentially dependent on the electronic noise present in the environment in which the detector is installed, it is of paramount importance to test it in a real-world application. We installed a prototype of the detector on the pELBE positron beamline of the HZDR accelerator complex. This is a pulsed positron beam produced by impinging the ELBE primary electron beam onto a tungsten target and then accelerating the re-emitted positrons to an energy of \SI{2}{\kilo eV}~\cite{EPOS}. Even though the beam is pulsed, its repetition frequency (\SI{1.625}{MHz}) is much higher than the detector bandwidth so that it can be treated, to all effects, as a continuous beam. The beam is transported from the production target to an experimental area where positron annihilation lifetime spectroscopy can be performed. We have installed our detector \SI{2.5}{m} upstream of the experimental chamber; closely downstream from a gate valve, a \SI{20}{\milli m} restriction in the beam tube and a pair of correction coils that allow steering the beam (henceforth called the horizontal and vertical coil, with the names corresponding to their actuation direction with respect to gravity). The detector was installed with the manipulator actuating horizontally, as in the orientation of figure \ref{fig:Assembly}c. The position at which we installed the detector makes it possible to operate it during sample changes, this allowed us to test the detector in a parasitic mode.

\subsection{Stability}
\label{subsec:Stability}

\begin{figure}
    \centering
    \includegraphics[width=0.90\linewidth]{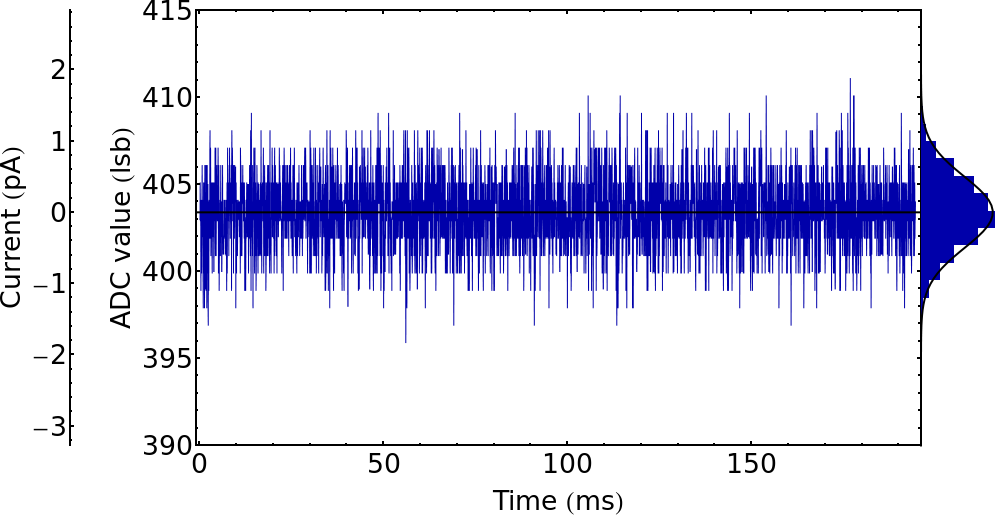}
    \caption{Noise level at the ADC, assessed by acquiring 4000 samples at the maximum rate permitted by the firmware. The distribution is normal, with a width of about \SI{2}{lsb}, typical for all four elements.}
    \label{fig:DitheringSpread}
\end{figure}

Multiple acquisitions from the same ADC produces samples that are distributed as a normal distribution with variance $\SI{1.9 \pm 0.2}{lsb}$ (see figure \ref{fig:DitheringSpread}).
Repeated sampling over a longer time interval show a slow drift of the average with a rate of $\SI[print-unity-mantissa=false]{1e-4}{lsb/s}$. These two characteristics combined allow us to increase both the precision and resolution of the measurement by performing repeated measurements in short succession. With this method the current resolution $\sigma_I$ will be:

\begin{equation}
    \sigma_I ~=~ \frac{1.9 \, k_\text{lsb}}{\sqrt{N}} 
    ~=~ \frac{k_t}{\sqrt{t}} \label{DitheringScaling}
\end{equation}
where $N$ is the number of consecutive measurements, $t$ is the measurement time and the conversion factors are $k_\text{lsb} = \SI{0.24}{\pico A}$ and $k_t = \SI{1.9} k_\text{lsb} / \sqrt{\text{sampling rate}}$, which is equal to \SI{3.2}{\femto A \sqrt{s}} when a single electrode is being sampled or to \SI{5.9}{\femto A \sqrt{s}} when all four electrodes are being read out. The longest sampling time when reading out all electrodes with the current firmware is obtained by sampling each of them 64000 times, which takes \SI{12.4}{s}; this measurement allows, according to eq.~\ref{DitheringScaling} the determination of the analog frontend output to $\SI{7.5e-3}{lsb}$ or \SI{1.8}{\femto A}. During the measurement we can expect the drift to contribute an additional \SI{0.12}{\femto A} of bias, making equation \ref{DitheringScaling} a suitable estimation of the digitization precision.

\begin{figure}
    \centering
    \includegraphics[width=0.90\linewidth]{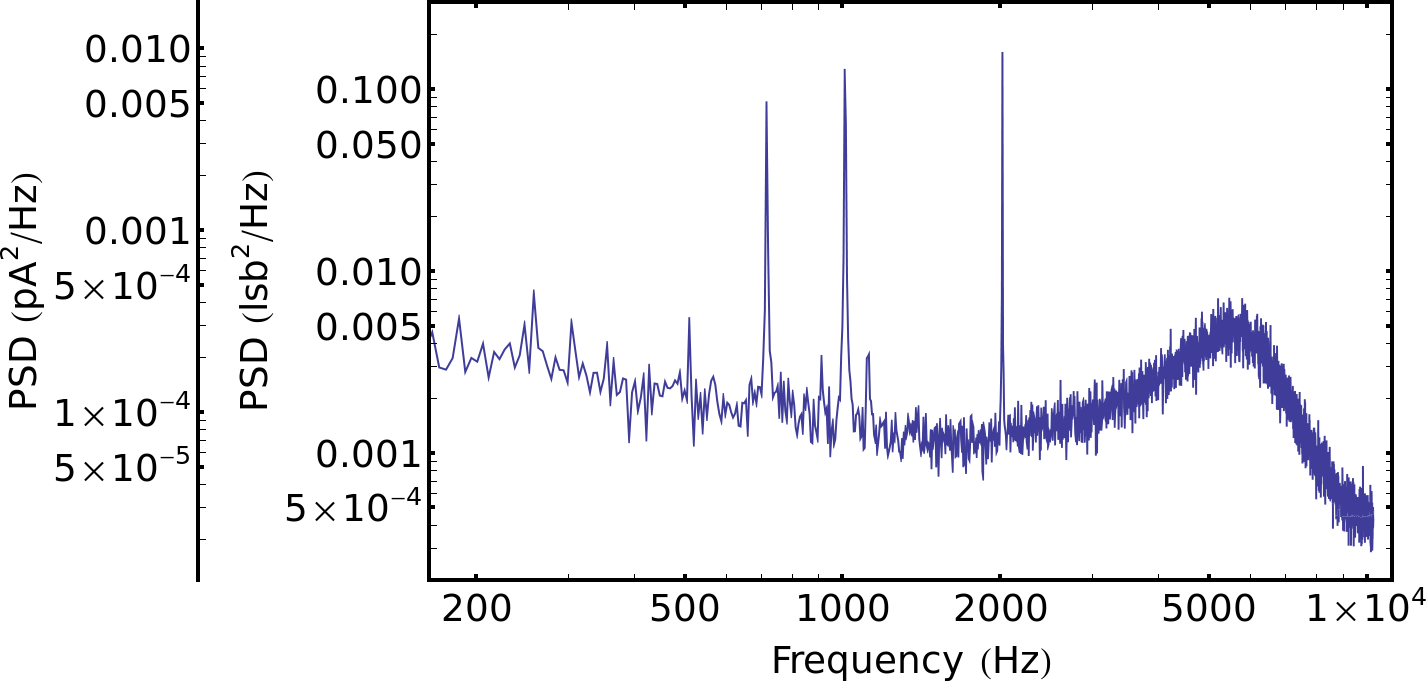}
    \caption{Power density spectrum of the noise recorded by the ADCs. The plot is an average of 30 consecutive runs, each 4000 samples long. We can see three spikes around \SI{720}{Hz}, \SI{1}{kHz} and \SI{2}{kHz}.\textbf{}}
    \label{fig:NoiseSpectrum}
\end{figure}

Spectral analysis of the noise (see figure \ref{fig:NoiseSpectrum}) shows a continuous spectrum with three main spikes at \SI{720}{Hz}, \SI{1}{kHz} and \SI{2}{kHz} present in the spectra from all electrodes. Considered the frequencies and the context in which the measurement was performed, we attribute these to the feedback loop of the power supplies powering the guidance coils installed around the beam line creating ripples at these specific frequencies.

\begin{figure}
    \centering
    \includegraphics[width=0.90\linewidth]{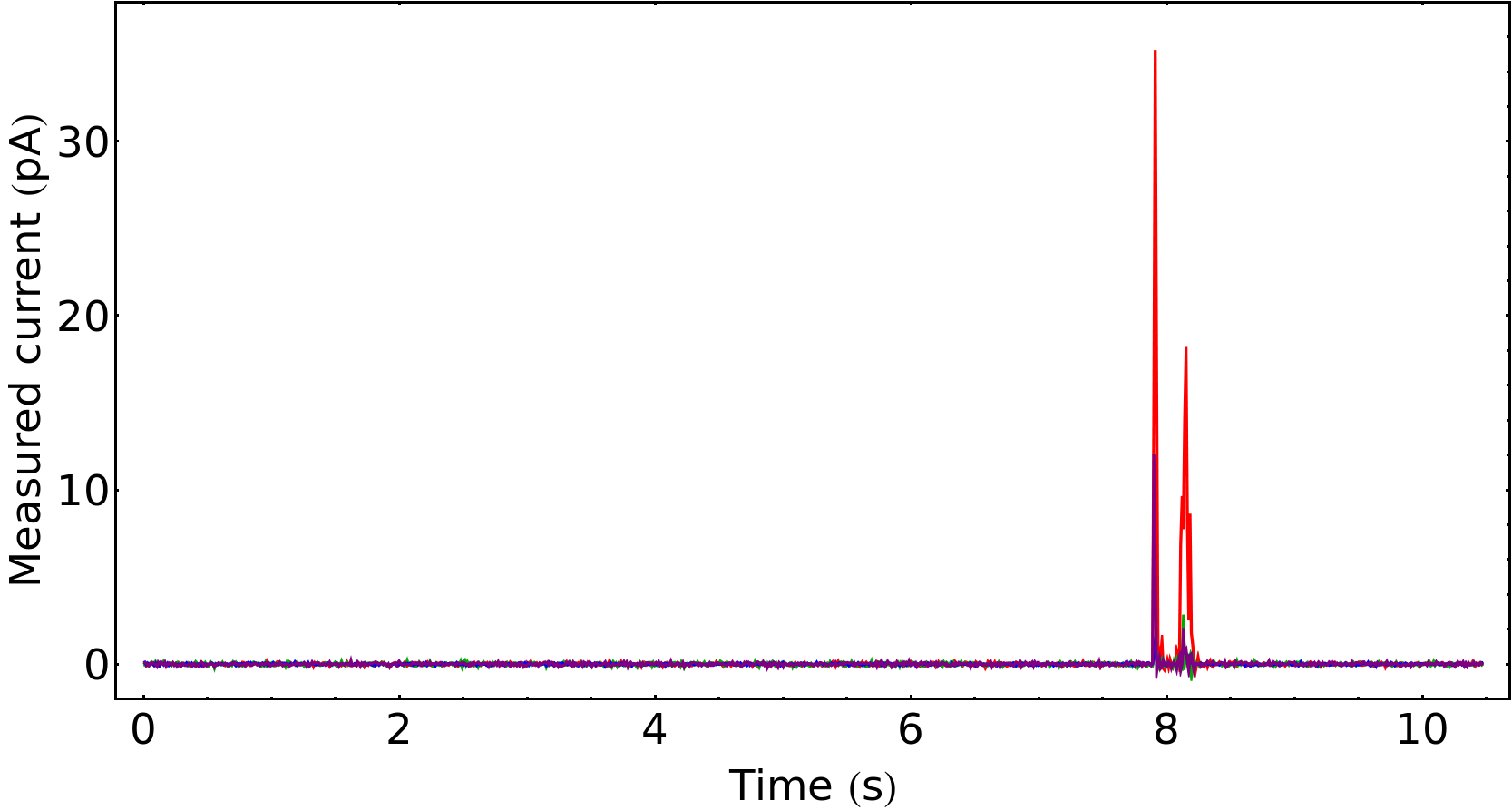}
    \caption{During the course of a long acquisition, the solenoid-mediated actuation of a gate valve creates a noise spike of more than 50 standard deviations at \SI{7.8}{s}, visible on all four detectors. These occurrences are easily identified and have to be removed from the data before averaging.}
    \label{fig:SingularEvent}
\end{figure}

As we have experienced with our measurements, a high vacuum chamber serves as an excellent Faraday cage against interference form electronics installed nearby the apparatus. The switching of devices placed in the chamber is liable, though, to generate electronic interference with the extremely sensitive preamplifier stage of the detector, we found this to be the case whenever the sample changing procedure required the actuation of gate valves, whose governors are solenoid-initiated. Whenever such an actuation occurs its interference is clearly visible in the recorded data (see figure \ref{fig:SingularEvent}) and can be cut out from the recording. We have filtered these singular events by removing every data point at more than 5 standard deviations from the mean, along with the 8 points preceding and following it, then repeating the procedure until a fixed point is reached.

\begin{figure}
    \centering
    \includegraphics[width=0.90\linewidth]{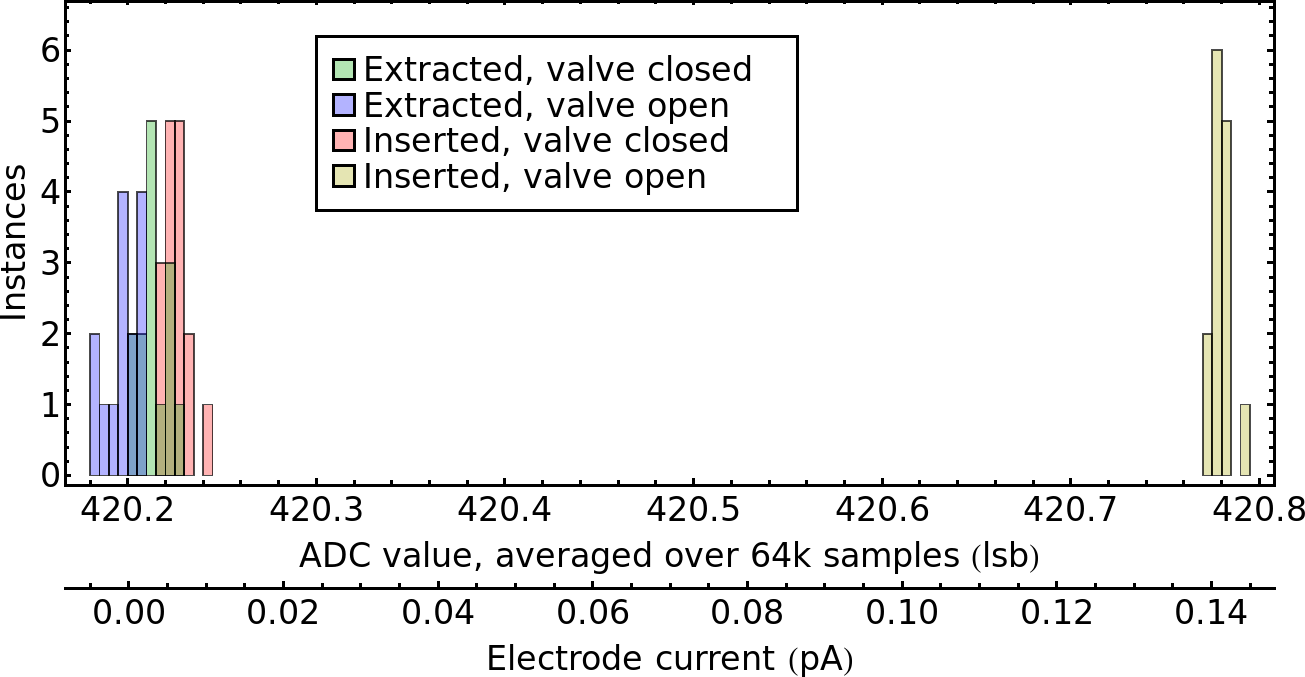}
    \caption{Assessment of the stability of the measurement performed by alternating the measurement of four different situations. During the assessment the detector is repeatedly inserted into the beam path and then extracted out, while a gate valve positioned closely in front of the detector location is alternatively opened and closed. For clarity, results from only one of the four electrodes are reported. We can distinctly see the signal produced by the positron beam when the detector is inserted and the valve is open.}
    \label{fig:Stability}
\end{figure}

Regardless of the precision of the digitization, the detector sensitivity is ultimately limited  by the precision of the analog frontend, which due to the design employed, we do expect to be much higher than what would be predicted from the minimum guaranteed manufacturer specifications. On top of that, it is not unreasonable to posit that by inserting or extracting the detector to and from the beam path the leakage current could be altered, due to the assembly shifting around, thus invalidating a reference measurement. We assessed both this sources of uncertainty by measuring 64000 samples from each electrode for 58 consecutive times while alternating having the detector inserted in the beam path or not and the gate valve being open or closed. The result shows great reproducibility (see figure \ref{fig:Stability}), with each condition being distributed around its own mean value. The average standard deviation from each group of measurements is $\SI{6.9e-3}{lsb}$ which is reasonably close to the $\SI{7.5e-3}{lsb}$ predicted by eq.~\ref{DitheringScaling}, hence indicating that the analog frontend is likely not limiting the precision of the measurement.

\subsection{Profiling}
\label{subsec:profiling}

\begin{figure}
    \centering
    \includegraphics[width=0.85\linewidth]{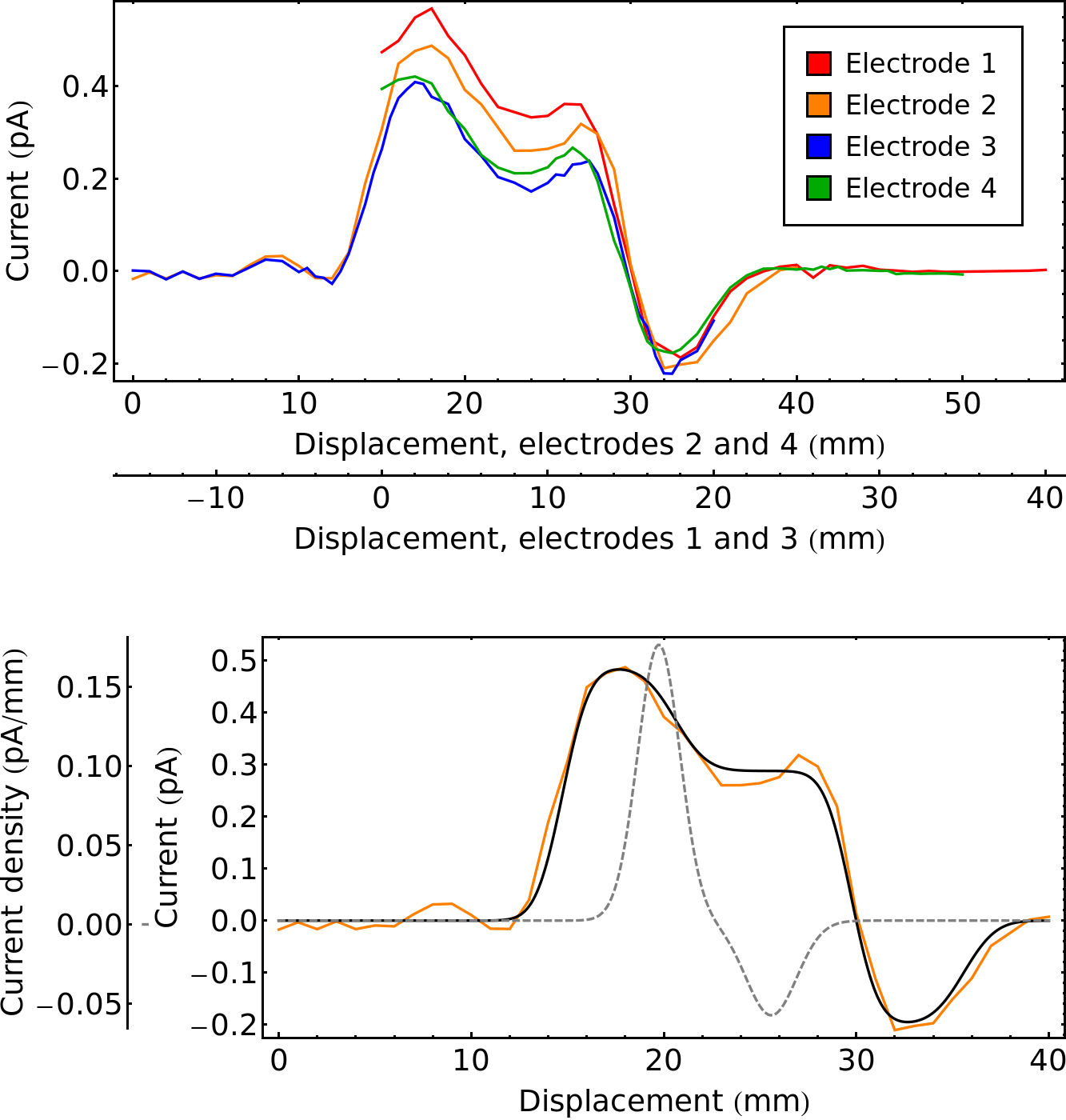}
    \caption{Top: Profiling of the beam done by progressively extracting the detector from the chamber, the horizontal scale has been shifted to account for the different position of the electrodes in the detector. Bottom: the profile recorded by electrode 2 (orange) and the minimum $\chi^2$ to it given by a model consisting of two gaussian-shaped beams convolved with a \SI{15}{\milli m}-wide rectangular window. In dashed gray, the two gaussian shaped profiles before convolution.}
    \label{fig:Profiling}
\end{figure}

It is possible to use this detector to measure the beam profile. The most natural way to proceed is to use the manipulator to partially extract the detector and measure the beam intensity as a function of the detector position while using the vertical deflection coil to center it vertically with the electrodes. We profiled the beam using electrodes 1 and 2 while applying \SI{-5}{A} of vertical steering current and with electrodes 3 and 4 while applying a current of \SI{10}{A}: by applying these currents the beam was roughly aligned with the center of each pair of electrodes during the scan. Figure~\ref{fig:Profiling} (top) shows the resulting profiles. To further test the stability of the apparatus the scan has been executed by first measuring in succession all points in which the manipulator was moved by an integer amount of millimeters (that is every second point in the scan) and then by measuring in reverse order the half integer displacements; the results from the two scans interleave properly, indicating once more an absence of significant drift in the apparatus. The profile recorded by all detectors is identical in shape, while differing slightly in amplitude. The difference in amplitude is mostly due to the beam being clipped differently by the detector edges in the direction orthogonal to the manipulator actuation and in part due to a difference in the absolute calibration of the detector scale, which is determined by the actual resistance value of the gigaohm resistor installed in the preamplifier.

In order to interpret the shape of the profile we should first of all restrict ourselves to consider its portion comprised in the \SI{15}{\milli m} wide stripe swept by the electrodes and then project the beam intensity along the manipulator's direction of motion. As we extract the detector from the beam path, we expect the electrodes to integrate the projected beam intensity over a window with the same width of the electrodes themselves. If the beam is gaussian-shaped and smaller than the electrode width, this will result in a \SI{15}{\milli m}-wide plateau flanked by error-function-shaped inclines connecting it to the background level. Instead our profile exhibits a plateau that dips midway through like the center of a fedora hat to then decline below the background level after the end of the \SI{15}{\milli m} window and returns to zero about \SI{8}{\milli m} later. This profile is easily explained by the presence of two beams inside the chamber, spaced \SI{6}{\milli m} apart and having opposite charge. Figure~\ref{fig:Profiling} (bottom) shows a least $\chi^2$ fit to the profile recorded by electrode 2 done using two gaussian profiles convolved with a \SI{15}{\milli m} wide rectangular window, showing a good adherence of the model to the experimental data.

As will be confirmed by the measurements shown in section \ref{sec:Twodee}, the negative peak is most likely an electron beam being transported alongside the positrons in the beamline. Its presence is not surprising, as the generation of positrons through pair production is expected to generate a shower of electrons with a wide spectrum of energies alongside the positrons, a phenomenon already observed in the NEPOMUC beamline~\cite{HugenschmidtBifurcatedBeam}. The portion of the produced electrons that possess enough energy to overcome the potential barrier of the acceleration voltage used to form the positron beam will enter the beamline alongside it. Since adiabatic transport is not sensitive to the sign of the charge of the transported particles, the electrons will travel alongside the positrons, with only the non-adiabatic components of the transport working to separate the two.

\subsection{Two-dimensional beam scan}
\label{sec:Twodee}

A different method of profiling the beam is to use the correction coils installed on the beamline in proximity to the detector to deflect the beam. We can determine the control authority of the coil that deflects the beam in the direction of the manipulator's actuation by comparing the profiling of the beam obtained from the mechanical extraction of the detector with the profile obtained by actuating the deflection coils. We employed 6 different scans by using different electrodes and vertical coil currents to determine the horizontal coil control authority to be $\SI{1.35 \pm 0.05}{\milli m \per A}$. Since the horizontal and vertical displacement coils are built identically, we can apply the same value to the vertical displacement.

\begin{figure*}
    \centering
    \includegraphics[width=0.95\linewidth]{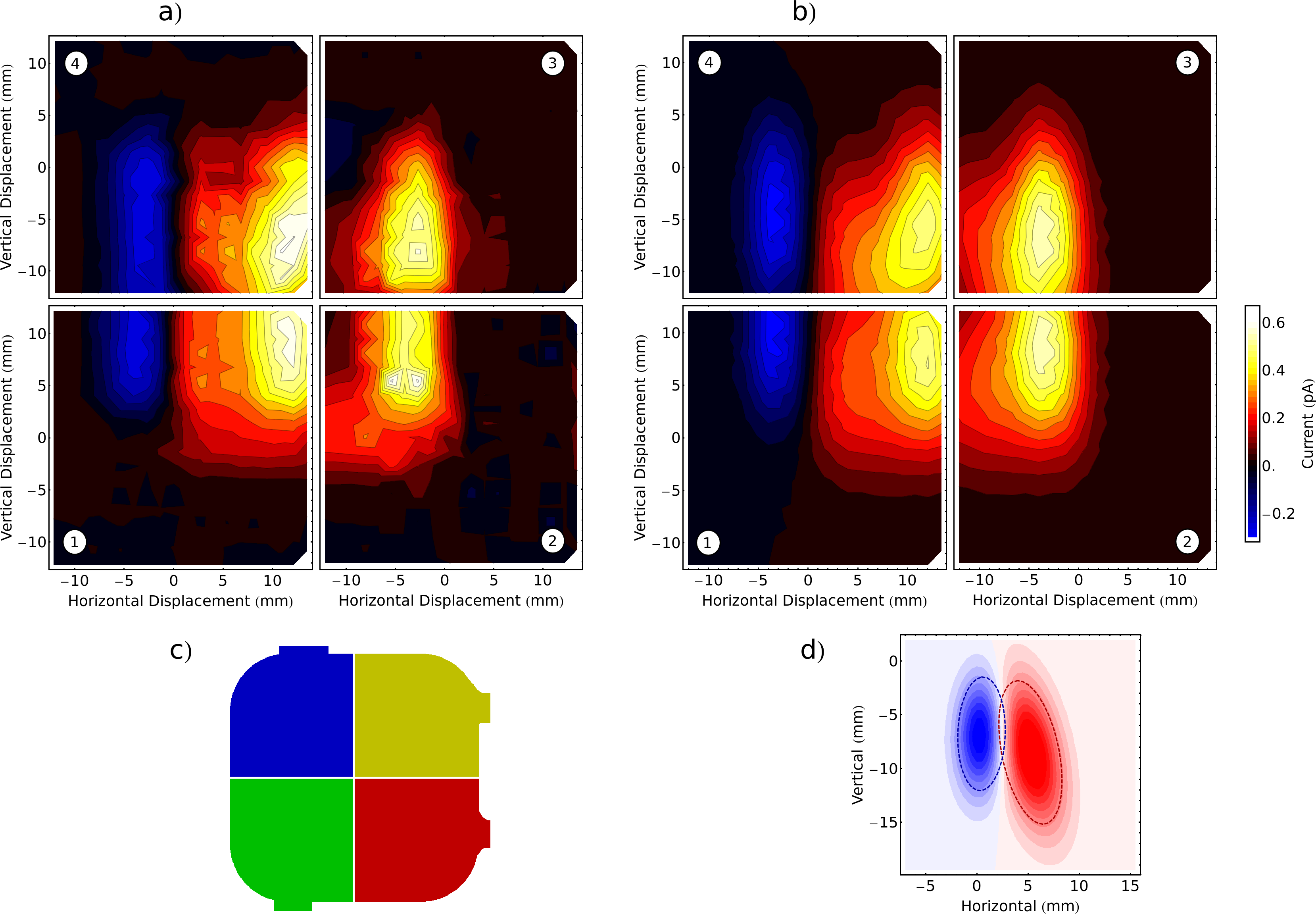}
    \caption{a) Two-dimensional scan of the beam for different settings of the deflection coils (165 points) showing the readout currents of the four Faraday cups. b) Minimum $\chi^2$ fit to the data by using a 12 parameters model with two gaussian beams. c) Shape of the electrodes employed for the numerical integration of the beam intensity. d) Reconstructed beam shape. }
    \label{fig:Quadrants}
\end{figure*}

Performing the two dimensional scan with all four electrodes is expected to sample a similar map of intensities, shifted horizontally and vertically by \SI{15}{\milli m} due to the different placement of the electrodes onto the detector head. Figure~\ref{fig:Quadrants}a shows a two dimensional scan of the beam using all detectors. We can see that we can easily stitch together the four scans into a coherent image by superimposing the portion of the plots neighboring the center of the quadrant. As anticipated in section \ref{subsec:profiling} the negative peak is caused by the presence of an electron beam traveling alongside the positrons. Similarly to section \ref{subsec:profiling} we can fit the entire scan with a beam model convoluted with the proper mask. This time we used the actual shape of the electrodes (shown in figure \ref{fig:Quadrants}c) to convolve the beam model. We defined a 12-parameter beam model consisting in the sum of two bi-dimensional gaussian profiles, each having as free parameters the peak coordinates, the peak amplitude, the width along two orthogonal axes and the angular orientation of the axes. Figure~\ref{fig:Quadrants}d shows the beam profile resulting from the fit operation, while figure \ref{fig:Quadrants}b shows the result of the scan as predicted by this model, which closely resembles the profile measured in \ref{fig:Quadrants}.

The reconstruction indicates the presence of two elliptic beams (aspect ratio $\approx 2.5$) of opposite charge having their major axes mostly aligned along the slightly tilted with respect to each other and the measurement frame of reference. Positrons are produced at ELBE by implanting a high energy electron beam onto a tungsten target and are then transported to the detector installation point by adiabatic transport through three bends roughly aligned with the $y$ axis of the plot in figure \ref{fig:Quadrants}d. The elliptical shape of the beam is expected, as the tungsten is installed with a $45^\circ$ inclination with respect to the ELBE primary beam~\cite{EPOS}, although the high aspect ratio of the reconstructed beam spots hints at the beam being already elliptical in shape before impinging on the conversion target. Adiabatic transport along a bent tube is known to cause the beam to drift in the direction of the rotation axis; this effect is proportional to the inverse of the curvature radius~\cite{NORTHROP1961} causing a shear deformation of the transported beam. The direction in which the shearing is expected to take place depends on the sign of the charge of the particles constituting the beam, which is precisely that we observe in the fit results. Despite both beam spots fitting well in this model, we need to restrict our considerations only to the positron beam, as the elliptic shape of the electron beam could also be explained as the result of masking by a beamline restriction of a larger, more diffused, electron beam. We have two reasons to conclude that this is indeed most likely the case: firstly that the presence of a diffused electron beam had been indeed observed in the past at ELBE and reported in the experiment logbooks, secondly the fact that the relative positioning of the peaks is such that if the positron beam (whose transport has been subject of optimization) is reasonably centered in the aperture then the edge of the opening would be tangential to the observed electron beam.

If we na\"ively integrate the fit result we find the positron beam intensity to be $\SI{622 \pm 6}{\femto A}$ and the electron beam intensity to be $\SI{77 \pm 0.7}{\femto A}$, with the error being mostly due to the uncertainty over the exact value of the gigaohm resistor. This values do not correspond to the actual beam intensity, though, as we can expect about 20\% of the implanted positrons to be backscattered~\cite{ColemannBackscattering, AersBackscattering, GhoshBackscattering, NishimuraSecondaries} and the emission of an average of 1.5 secondary electrons per positron implanted~\cite{NishimuraSecondaries}, resulting in a positron beam intensity of $\SI{461 \pm 55}{\femto A}$ making the calibration of this correcting factors the main source of uncertainty in the detector. This is an unavoidable limitation of the present design which, however, does not impact the main aim of the detector since only relative measurements are required to determine whether the beam is centered onto the beamline.

\section{Expected performance for beam alignment}

We should consider a few possible applications of the detector described in this paper. For simplicity we will ignore the signal amplification effect given by the emission of secondary electrons and provide a lower bound to the detector sensitivity; in reality the detector will perform at least as well as we are predicting here.

The most intense low energy positron beam available today is the NEPOMUC primary beam. With an intensity of $\SI{1.14e9}{e^+/\second}$ (\SI{183}{\pico A}) and a diameter of \SI{9.3}{\milli m}~\cite{HugenschmidtPrimaryBeam,Dickmann2020}, we could consider this beam to be well centered when it travels within \SI{1}{\milli m} of the beam tube axis. Using the framework of Manojlovic~\cite{manojlovic2011} we can determine that it is necessary to resolve a difference of \SI{31}{\pico A} to achieve this. According to equation \ref{DitheringScaling} this is doable with the readout of a single ADC sample; therefore, excluding time extrapolation techniques, the maximum speed at which we can assess whether the beam is centered is \SI{6}{\milli s} as determined, instead, by the low pass filter built into the analog frontend. In a beam optimization scenario this time has to be doubled to \SI{12}{\milli s} to account for data transmission rates and USB latency. The same holds true for the NEPOMUC beam after remoderation: the remoderated NEPOMUC positron beam has an intensity of $\SI{5e7}{e^+/\second}$ (\SI{8}{\pico A}) and a diameter of \SI{1.85}{\milli m}~\cite{Dickmann2020}, in this instance determining it to be off-center by \SI{1}{\milli m} requires resolving a difference of \SI{5.7}{\pico A} which also does not require dithering.

We have determined the pELBE beam to have an intensity of $\SI{2.87e6}{e^+ / s}$ (\SI{461}{\femto A}) and a width of \SI{11.25}{\milli m} FWHM vertically and \SI{4.5}{\milli m} horizontally. Determining its vertical centering requires resolving a difference of \SI{67}{\femto A}, which requires dithering with the averaging of 100 samples from each electrode, resulting in an acquisition time of \SI{1.6}{\milli s}; after transmission delay is added this minimally lengthens the measurement time needed for more intense beams.

The source-based beam installed in Garching has an intensity of \SI{6.0e4}{e^+/s} (\SI{9.6}{\femto A}) with a diameter below \SI{5}{\milli m}~\cite{LabBeam}, which requires a resolution of \SI{2.11}{\femto A} to determine misalignment. This can be achieved through dithering by sampling for \SI{2.1}{s}. Differently from the other scenarios considered here, this measurement time can potentially be relevant in terms of limiting the speed at which a beam optimization procedure can be performed and could benefit from the improvement of the digitization stage of this design, as we will propose later.

\section{Conclusion and Outlook}

We have shown that the current setup can quickly resolve currents with sub-picoampere resolution and can reach a peak precision of \SI{1.8}{\femto A} by using longer acquisition times. It is fast and sensitive enough to be able to resolve the misalignment of exotic particle beams and allow beam optimization through very fast iterations in a variety of cases. The implementation of the detector is compact and inexpensive and can thus be widely deployed at facilities making use of exotic particle beams.

When considering the specific case of source-based positron beams the measurement time becomes unfortunately significantly longer, due to the dithering required to assess the output of the analog frontend with sufficient precision. We can address this issue with three improvements:

\begin{enumerate}
    \item Increasing the bit depth of the ADC
    \item Increasing the speed of the ADC
    \item Increasing the gain of the differential amplifier feeding into the ADC
\end{enumerate}

The implementation of the latter point requires the introduction of a way to actively introduce a bias to the differential amplifier input to prevent the output from reaching the limit of its range due to the leakage current unavoidably present in the system. We have already reviewed the design to include all these improvements, although the revised design still has to be tested.

\begin{figure}
    \centering
    \includegraphics[width=0.90\linewidth]{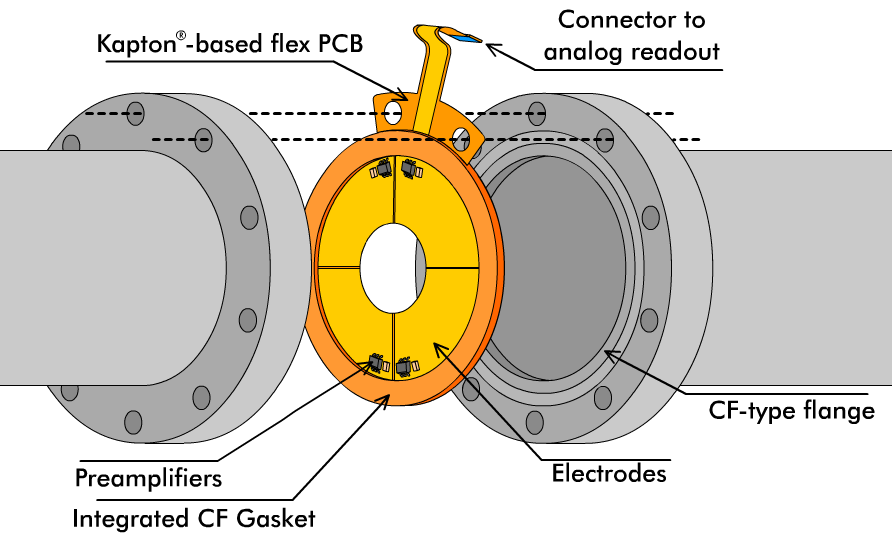}
    \caption{Simplified mechanical design of a Faraday cup detector of off-centering of the particle beam. }
    \label{fig:NovelDesign}
\end{figure}

Another potential improvement for the present sensor is to realize it not as a beam stopper segmented into four quadrants, but instead as a membrane with a hole in the center, with the four electrodes surrounding it (see figure \ref{fig:NovelDesign}). In this variant the sensor would be able to detect the beam being significantly off-center and would provide information about the direction in which the beam was lost. The main advantage of this configuration is that it requires no moving parts, reducing by about 80\% the material costs needed to realize it. We believe that this specific configuration could be realized in the form of a custom kapton-based ring holding the detector membrane and serving both as a vacuum gasket and a passthrough for the readout lines, which has the combined advantage of further halving the construction costs and requiring only minimal modifications to be installed in preexisting apparatuses.

We will be installing multiple instances of the sensor as designed for this test along the existing NEPOMUC beamline and its envisioned extension into the EAST Hall, and will continue the development of novel variants of this design.

\section*{Acknowledgements}

We would like to thank Dr. Michael B\"ohmer and the electronics laboratory of the physics department of the Technical University of Munich for the assistance provided with this project and Dr. Davide Orsucci for reviewing and improving the drafts of this paper.

\bibliographystyle{elsarticle-num} 
\bibliography{cas-refs}

\vfill

\end{document}